\documentclass[draft,grl]{agutex}
\usepackage{lineno}
\usepackage{graphicx}
\authorrunninghead{SHUKLA}
\titlerunninghead{FULL-SCALE SIMULATION STUDY}
\authoraddr{P. K. Shukla, International Centre for Advanced Studies in Physical Sciences \&
Institute for Theoretical Physics, Ruhr University Bochum, 44780 Bochum, Germany (profshukla@yahoo.de)}

\linenumbers*[1]
\begin{document}
%% ------------------------------------------------------------------------ %%
%
%  ENABLE IMAGE DISPLAY WHILE USING DRAFT MODE
%
%% ------------------------------------------------------------------------ %%
%
% Uncomment the following code (as well as \usepackage{graphicx} above)
% if you need to include images in draft mode
%\setkeys{Gin}{draft=false}
%
% PLEASE NOTE: WHEN YOU SUBMIT YOUR LATEX FILE TO GEMS, COMMENT OUT ANY COMMANDS
% THAT INCLUDE GRAPHICS.
% (See FIGURES section near the end of the file)
%
%% ------------------------------------------------------------------------ %%
%
%  TITLE
%
%% ------------------------------------------------------------------------ %%
\title{Alfv\'enic tornadoes}
%
% e.g., \title{Terrestrial Ring Current:
% Origin, Formation and Decay $\alpha\beta\Gamma\Delta$}
%% ------------------------------------------------------------------------ %%
%
%  AUTHORS AND AFFILIATIONS - 3 methods
%
%% ------------------------------------------------------------------------ %%
% Method 1 (for all journals, except Reviews of Geophysics, which
% should use method 3):
% For three or fewer author/affiliation blocks, use \author{} and \affil{}

\author{P. K. Shukla}
\affil{International Centre for Advanced Studies in Physical Sciences \& Institute for Theoretical Physics,
Ruhr University Bochum, D-44780 Bochum, Germany}

\affil{Department of Mechanical and Aerospace Engineering \& Center for Energy Research, University of California San Diego,
La Jolla, CA 92093, USA}

\begin{abstract}
It is shown that a three-dimensional (3D) modified-kinetic Alfv\'en waves (m-KAWs) can propagate
in the form of  Alfv\'enic tornadoes characterized by plasma density whirls or   magnetic flux ropes
carrying orbital angular momentum (OAM). By using the two fluid model, together with  Amp\`ere's law,
we derive the  wave equation for a 3D m-KAWs in a magnetoplasma with $m_e/m_i \ll \beta \ll 1$,
where $m_e$ $(m_i)$ is the electron (ion) mass, $\beta =4 \pi n_0  k_B (T_e + T_i)/B_0^2$, $n_0$
the unperturbed plasma number density, $k_B$ the Boltzmann constant, $T_e (T_e)$ the electron (ion)
temperature,  and $B_0$ the strength of the ambient magnetic field. The 3D m-KAW equation admits
solutions in the form of a Laguerre-Gauss (LG) Alfv\'enic vortex beam or Alfv\'enic tornadoes
with plasma density whirls that support the dynamics of Alfv\'en magnetic flux ropes.
\end{abstract}
%% ------------------------------------------------------------------------ %%
%
%  BEGIN ARTICLE
%
%% ------------------------------------------------------------------------ %%
% The body of the article must start with a \begin{article} command,
% and an \end{article} command must be placed at the end of the file,
% before \end{document}.
%
% If using draft mode \end{article} must follow the references section.
\begin{article}
%% ------------------------------------------------------------------------ %%
%
%  TEXT
%
%% ------------------------------------------------------------------------ %%
\section{Introduction}

The discovery of the Alfv\'en wave \citet{Alfven} had significantly changed the landscape of collective phenomena in
magnetized plasmas that frequently occur in  interstellar media \citet{Dawson} and in large laboratory  plasma
devices \citep{Walter99,Kletzing10, Walter11}.  In the  Alfv\'en wave, the restoring force comes from the pressure
of the magnetic fields,  and the ion mass provides the inertia.  The dispersion relation for the low-frequency
(in comparison with  the ion gyrofrequency) Alfv\'en wave is deduced from the magnetohydrodynamic (MHD)
equations \citep{Alfven,Cramer}, which do not account for the wave dispersion effects. The
dispersion \citep{Stefant,Brodin,Morales94,Morales97,Shukla99,Dastgeer,Jim} to the Alfv\'en wave comes
from the finite frequency  ($\omega/\omega_{ci}$) and the magnetic field-aligned electron inertial force
in cold magnetoplasmas,  as well as from  the finite ion gyroradius effect and the gradient of the electron
pressure perturbation in  a warm magnetoplasma,  Due to non-ideal effects \cite{Brodin,Morales94,Morales97,Shukla99,Jim},
Alfv\'en waves thus couple to fast and slow magnetoacoustic waves, the inertial and kinetic Alfv\'en
waves \cite{Stefant,Morales94,Morales97,Dastgeer,Jim} in a uniform magnetoplasma. The magnetic field-aligned
phase speed ($\omega/k_z)$ of the inertial (kinetic) Alfv\'en wave  is much larger (smaller) than the electron
thermal speed.  Accordingly, the inertial (kinetic) Alfv\'en wave arises in a plasma with
$\beta \ll m_e/m_i$ ($m_e/m_i \ll \beta \ll 1)$.  The dispersive shear Alfv\'en wave plays a significant
role in the auroral \cite{Chris08a,Chris08b}, magnetospheric \cite{David05},  solar \citet{Tripathi10,Salem12},
astrophysical  and laboratory \citep{Walter11,Walter12} magnetoplasmas with regard to acceleration and
heating of  the plasma  particles \cite{Shukla98}, reconnection of magnetic field lines, wave-wave and
wave-particle  interactions \citep{Hasegawa76,Carter06,Shukla07}, and the formation of nonlinear Alfv\'enic
structures \citep{David05,Sta03,Shukla12a}.  Specifically, it is likely that the dispersive Alfv\'en waves
power the auroral activities \citet{Chris08a,Chris08b} and the solar  coronal heating \citep{Tom,Science09,Nature11},
as well as may be responsible for the formation of Alfv\'enic tornadoes \citet{Nature12} which can transport
magneto-convective energy from one region to another in solar plasmas.

In this Brief Report, it is shown that a 3D m-KAW can propagate as a twisted vortex beam in a plasma with
$m_e/m_i \ll \beta \ll 1$. The present work thus complements a recent investigation \cite{Shukla12b} that
focused on examining a 3D twisted inertial Alfv\'en wave. Twisted m-KAWs  can be related with
  magnetic  flux ropes or Alfv\'enic tornadoes of different scalesizes in observational data from laboratory
experiments \citet{Walter12}, and also from the solar plasmas \citet{Nature12,Salem12}.

\section{Mathematical description}

We consider a magnetized electron-ion plasma in the presence of low-frequency (in comparison with
the electron gyrofrequency $\omega_{ce} =eB_0/m_e c$, where $e$ is the magnitude of the electron charge,
$B_0$ the strength  of the external magnetic field $\hat {\bf z} B_0$, $m_e$ the  electron mass,
and $c$ the speed of light in vacuum,  $\hat {\bf z}$ the unit vector along the $z$-axis in a Cartesian
coordinate system) m-KAWs with the electric and  magnetic fields ${\bf E} = -\nabla \phi -(1/c)\hat {\bf z} \partial A_z/\partial t$
and  ${\bf B}_\perp =\nabla A_z \times \hat {\bf z}$, respectively, where $\phi$ and $A_z$ are the scalar and
magnetic field-aligned vector potentials, respectively. In the m-KAW fields, the electron density perturbation is obtained from

%1
\begin{equation}
 \frac{\partial n_{e1}} {\partial t}  +  \frac{c} {4\pi e} \frac{\partial \nabla_\perp^2 A_z } {\partial z} =0,
\end{equation}
where we have used the electron fluid velocity

%2
\begin{equation}
{\bf u}_e   \approx \frac{c}{B_0} \hat {\bf z} \times \nabla \phi - \frac{c k_B T_e}{B_0 n_0}
\hat {\bf z} \times \nabla n_{e1} +\hat {\bf z}\frac{c}{4\pi e n_0} \nabla_\perp^2 A_z,
\end{equation}
which uses Amp\` ere's law that relates $A_z$ and  the magnetic field-aligned electron fluid velocity
$u_{ez}$.  Here $n_{e1} (\ll n_0$) is a small electron density perturbation in the equilibrium density
$n_0$.

Since the parallel phase speed of the m-KAWs is much smaller than the electron thermal speed
$V_{Te}= (k_BT_e/m_e)^{1/2}$, one can neglect the electron inertia and obtain from the magnetic field aligned  electron momentum equation

%3
\begin{equation}
0 = e \frac{\partial \phi}{\partial z} + \frac{e}{c}\frac{\partial A_z}{\partial t}
- \frac{k_B T_e}{n_0}\frac{\partial n_{e1}}{\partial z},
\end{equation}
which dictates that the parallel electric force $-n_0 e E_z$ and $\partial p_1/\partial z$ are in balance,
 where the magnetic field-aligned electric field $E_z =-\partial \phi -c^{-1}\partial A_z/\partial t$ and
$p_1 =k_B T_e n_{e1}$ is the electron pressure perturbation.

We can now eliminate $A_z$ from Eqs. (1) by using (3), obtaining

%4
\begin{equation}
\left(\frac{\partial^2}{\partial t^2} -c^2 \lambda_{De}^2 \frac{\partial^2 \nabla_\perp^2}
{\partial z^2} \right) n_{e1} -\frac{c^2}{4\pi e} \frac{\partial^2 \nabla_\perp^2 \phi} {\partial z^2} =0,
\end{equation}
where $\lambda_{De} = (k_B T_e/4\pi n_0 e^2)^{1/2}$ is the electron Debye radius.

The perpendicular (to $\hat {\bf z}$) component of the ion fluid velocity ${\bf u}_{i \perp}$ is determined from

%5
\begin{equation}
\left(\frac{\partial^2}{\partial t^2} + \omega_{ci}^2 \right) {\bf u}_{i \perp}
= \frac{c \omega_{ci}^2}  {B_0} \hat {\bf z} \times \nabla \phi
-\frac{c \omega_{ci}}{B_0} \frac{\partial \nabla_\perp \phi}{\partial t},
\end{equation}
which is obtained by manipulating the perpendicular component of the ion momentum equation, where $\omega_{ci} =e B_0/m_i c$
is  the ion gyrofrequency and $m_i$ the ion mass. We have assumed that the m-KAW phase speed is much larger than the ion thermal
speed, and therefore neglected in Eq. (5) the contribution of the ion pressure gradient. The ions are assumed to be confined
in a two-dimensional ($x-y$) plane perpendicular to $\hat {\bf z}$. The magnetic field-aligned ion dynamics is unimportant
for the m-KAW s, since the phase speed of the latter is much larger than the ion-sound speed.

From the linearized ion continuity equation and Eq. (5), we readily obtain

%6
\begin{equation}
\left(\frac{\partial^2} {\partial t^2} + \omega_{ci}^2 \right) n_{i1}
-\frac{n_0 c \omega_{ci}} {B_0} \nabla_\perp^2\phi = 0,
\end{equation}
where $n_{i1} (\ll n_0)$ is a small ion number density perturbation.

Invoking the quasi-neutrality condition $n_{e1} =n_{i1} =n_1$, we eliminate $n_{i1}$ from Eq. (6)  by using Eq. (5),
obtaining the wave equation for a 3D m-KAW after elementary calculation

%7
\begin{equation}
\left[\frac{\partial^2} {\partial t^2} - C_A^2 \rho_s^2 \frac{\partial^2 \nabla_\perp^2} {\partial z^2}
- \frac{C_A^2} {\omega_{ci}^2} \left(\frac{\partial^2}{\partial t^2} + \omega_{ci}^2\right)
\frac{\partial^2}{\partial z^2} \right] n_1 = 0,
\end{equation}
where $C_A = B_0/\sqrt{4 \pi n_0 m_i}$ is the Alfv\'en speed, and $\rho_s = C_s/\omega_{ci}$ is the
sound gyroradius, with $C_s =(k_B T_e/m_i)^{1/2}$ being the ion-sound speed. The effect of the ion temperature
can  be incorporated by re-defining $\rho_s = (C_s/\omega_{ci})(1+3T_i/T_e)^{1/2}$.

Within the framework of a plane-wave approximation, assuming that $n_1$ is
proportional to  $\exp(-i\omega t + i {\bf k} \cdot {\bf r})$, where $\omega$ and ${\bf k}
(={\bf k}_\perp + \hat {\bf z} k_z$) are the angular frequency and the wave vector, respectively,
we Fourier analyze (8) to obtain  the frequency spectra \citet{Shukla99} of the modified (by the
$\omega/\omega_{ci}$ effect) KAWs

%8
\begin{equation}
\omega^2 = \frac{k_z^2 V_A^2 (1  + k_\perp^2 \rho_s^2)}{1 +k_z^2 \lambda_i^2},
\end{equation}
where ${\bf k}_\perp$ and $k_z$ are the components of ${\bf k}$ across and along $\hat {\bf z}$, and
$\lambda_i =c/\omega_{pi}$ the ion inertial scale length, with $\omega_{pi} =(4\pi n_0e^2/m_i)^{1/2}$ being
the ion plasma frequency. The $k_z \lambda_i$-term in Eq. (8) comes from the
$\omega/\omega_{ci}$ effect.

Let us now study the property of a  twisted m-KAW.  We  seek a solution of Eq. (7) in the form

%9
\begin{equation}
n_1 = N  (r)\exp( i k_z z - i \omega_k t),
\end{equation}
where $N  (r)$ is a slowly varying function of $z$. Here $r =(x^2 + y^2)^{1/2}$ and $k$  is the
propagation wave number along the axial (the $z-$ axis) direction. By using Eq. (10) we can write
 Eq. (7)  in a paraxial approximation (viz. $|\partial^2 N /\partial z^2| \ll k_z^2 N$) as

%10
\begin{equation}
\left(2 i  \frac{\partial}{\partial Z} + \nabla_\perp^2\right)  N = 0,
\end{equation}
where $\omega_k = k_z V_A/(1+k_z^2 \lambda_i^2)^{1/2}$ is the angular frequency of the magnetic
field-aligned KAW.  Here $Z = (1+ k_z^2 \lambda_i^2) k_z z $ and $\nabla_\perp$ is in unit  of $\rho_s$.
 Furthermore, we have denoted the operator $\nabla_\perp^2 N=(1/r) (\partial/\partial r)
(r\partial N/\partial r) + (1/r^2)\partial^2 N/\partial \theta^2$, and introduced the cylindrical coordinates
with ${\bf r} =(r,\theta,z)$.

The solution of Eq. (10) can  be written as a superposition of Laguerre-Gaussian (LG) modes \citep{Allen92,Allen03,Tito09},
each of them representing a state of orbital angular momentum, characterized  by the quantum number $l$,  such that

%11
\begin{equation}
N =\sum_{pl} N_{pl} F_{pl}(r,Z) \exp(i l \theta),
\end{equation}
where the mode structure function is $ F_{pl}(r,z) = H_{pl} X^{|l|} L_p^{|l|}(X) \exp(-X/2$),
with $X=R^2/W^2(Z)$, $r=r/\rho_s$, and $W= 2\pi  (1+4\pi^2 \lambda_i^2/L_z^2) z/L_z$,
$L_z =2\pi/k_z$,  and $W(Z)/L_z$ representing the  normalized width of a twisted Alfv\'enic vortex beam.
The normalization factor $ H_{pl}$ and the associated Laguerre polynomial $L_p^{|l|}(X)$ are, respectively,

%12
\begin{equation}
H_{pl} = \frac{1}{2\sqrt{\pi}}\left[\frac{(l+p)!}{p!}\right]^{1/2},
\end{equation}
and

%13
\begin{equation}
L_p^{|l|}(X) = \frac{\exp(X)}{X p!} \frac{d^p}{dX^p}\left[X^{l+p} \exp(-X)\right],
\end{equation}
where $p$ and $l$ are the radial and angular mode numbers of the DSAW orbital angular momentum state.
In a special case with $l=0$ and $p=0$, we have a Gaussian beam.

The LG solutions, given by Eq. (11), describe the structure of a twisted 3D Alfv\'enic vortex beam carrying OAM.
In a twisted m-KAW vortex beam, the wavefront rotates around the beam's propagation direction (viz. the $z$-axis)
in a spiral that looks like  fusilli pasta (or a bit like a DNA double helix), creating a vortex and leading to
the m-KAW vortex (KAWV) beam with zero intensity at  its center. A twisted m-KAWV beam (or an Alfv\'en tornado)
can be created with the help of two oppositely  propagating 3D m-KAWs that are colliding in a magnetoplasma.
Twisting of 3D  m-KAWs occurs because different sections of the  wavefront bounce off different steps,
introducing a delay between the reflection of neighboring sections and, therefore, causing the wavefront
to be twisted due to an entanglement of the wavefronts.  Thus, due to angular symmetry, Noether theorem
guards OAM conservation  for a  m-KAWV beam that is accompanies the parallel electric field, sheared magnetic field,
and finite density perturbations.

\section{Summary and discussions}

In summary, we have shown that 3D m-KAWs in a uniform magnetoplasma can propagate in the form
of Alfv\'enic tornadoes which have axial and radial extents of the order of the ion inertial length
$c/\omega_{pi}$  and the ion sound radius $\rho_s$, respectively. The Alfv\'enic swirls  can be identified
as observational signatures of rapidly rotating magnetic field ropes \citep{Tripathi10,Walter12}
(or magnetic field tornadoes), which can provide an alternative mechanism for channelling electromagnetic energy
and heat fluxes from the surface  of the Sun into the corona \cite{Nature12} through Alfv\'enic tornadoes/whirls.
Furthermore, the present investigation  of twisted m-KAW vortex  beams can also be
useful  for diagnostic purposes, since the m-KAW beam frequency can be a fraction of the ion
gyrofrequency.

\acknowledgments
This research was partially  supported by the Deutsche Forschungsgemeinschaft (DFG), Bonn, through the
project SH21/3-2 of the  Research Unit 1048.

\end{article}
\end{document}